# Space-coiling metamaterials with double negativity and conical dispersion

Zixian Liang[1*], Tianhua Feng[1*], Shukin Lok[1*], Fu Liu[1], Kung Bo Ng[2], Chi Hou Chan[2], Jinjin Wang[3], Seunghoon Han[4**], Sangyoon Lee[4], and Jensen Li[1,2**],

[1]*Department of Physics and Materials Science, City University of Hong Kong, Tat Chee Avenue, Kowloon Tong, Kowloon, Hong Kong, China.*
[2]*State Key Laboratory of Millimeter Waves, City University of Hong Kong, Tat Chee Avenue, Kowloon, Hong Kong, China.*
[3]*Kuang-Chi Institute of Advanced Technology, Shen Zhen, Guangdong, China.*
[4]*Samsung Advanced Institute of Technology, Samsung Electronics, South Korea.*
[*]*These authors contribute equally to this work.*
[**]*email: jensen.li@cityu.edu.hk;shn.han@samsung.com*

Metamaterials are effectively homogeneous materials that display extraordinary dispersion. Negative index metamaterials, zero index metamaterials and extremely anisotropic metamaterials are just a few examples. Instead of using locally resonating elements that may cause undesirable absorption, there are huge efforts to seek alternative routes to obtain these unusual properties. Here, we experimentally demonstrate an alternative approach for constructing metamaterials with extreme dispersion by simply coiling up space with curled channels. Such a geometric approach has the advantage that the ratio between the wavelength and the lattice constant in achieving a negative or zero index is easily tunable. It allows us to construct for the first time an acoustic metamaterial with conical dispersion, leading to a clear demonstration of negative refraction from an acoustic metamaterial with airborne sound. We also design and realize a double-negative metamaterial for microwaves under the same principle.

Double negativity, zero index and extreme anisotropy are some of the unusual types of dispersion relations that can be realized using metamaterials, leading to applications including super-resolution imaging, invisibility cloaking and various transformation-based devices [1-22]. In particular, double-negative metamaterials, with separately tuned resonances in the two constitutive parameters, have been realized for electromagnetic waves [2-5], transmission-line circuits [6], and acoustic waves [16-21]. These developments have stimulated the search for alternative routes to achieve negative refraction including chirality, band folding, time reversal, kinetic inductance and Mie scattering, which may provide a wider bandwidth, a lower loss, or a larger scalability to bulk metamaterials [23-31].

One of these alternative routes is to use a photonic or phononic crystal to achieve band folding; based on this approach, a band with negative phase velocity emerges that enables negative refraction [29,30]. Although photonic and phononic crystals lack a valid effective medium description and the wave manipulation is difficult to go beyond the diffraction limit, they have been proven to be a close approximation to achieve various extreme wave phenomena originally found in metamaterials. Apart from negative refraction, sharp focusing or imaging, zeroth-order Bragg photonic gap, invisibility cloaking and recently conical dispersion have been demonstrated in this way [31-37]. These wide-ranging demonstrations not only provide an alternative route to employing locally resonating elements in conventional metamaterials but also an insight into the similarities between a photonic/phononic crystal and a metamaterial if we can reduce the band gap to low frequencies. Consequently, an effective medium can become meaningful and valid even at frequencies above the band gap. The extreme effective parameters at lower frequencies can allow us to manipulate waves in higher resolution and they can be further employed in the more general framework of transformation acoustics/optics in wave manipulations.

In this case, if we wanted to avoid using locally resonating elements exhibiting high loss, a low-frequency band gap would normally require constitutive materials with high refractive indices, which may not be readily found in nature. For example, in the domain of acoustics for airborne sound, common solids have indices lower than one, while materials with a much lower sound speed than that of air (e.g. silicone rubber) usually accompany substantial absorption loss. It is also difficult to have high-index materials with low losses in the optical domain. As a result, the lattice constant binds to the working frequencies above the first band gap. Interestingly, in the case of metamaterials that rely on locally resonating units, there is also usually a binding between the electrical size of the resonating units and the working frequency with negative indices. We show that this binding can be relaxed so that the ratio of the wavelength to the lattice constant can be easily tunable if we can yield a high refractive index by using a geometric route.

In this article, we experimentally demonstrate a geometric approach for constructing metamaterials with extreme dispersion by coiling up space. By delaying the propagation phase using curled channels to mimic an array of high-index elements, an acoustic metamaterial with conical dispersion and associated double negativity can be obtained at very low frequencies. As a result of our approach, we can demonstrate negative refraction from an acoustic metamaterial with airborne sound. Moreover, we also construct a double-negative metamaterial that works for microwaves under the same geometric principle.

In the current scheme, the mentioned high refractive index is mimicked by coiling up the wave propagation space through curled channels with substantial phase delays (Fig. 1a). The refractive index $n_{1D}$ of such a one-dimensional element is the ratio of the actual elapsed phase to the corresponding elapsed phase in a straight channel of the same physical length. Suppose we join these elements into a two-dimensional (2D) array. The result is that band-folding occurs at a low frequency and that the array has an effective medium description with extreme indices around the band-folding frequency at the zone center [38]. The wavelength ($\lambda$) with these extreme indices can therefore be easily tuned; further, it can be much larger than the lattice constant ($a$) by simply geometrically increasing the number of turns of the curled channels.

This space-coiling scheme applies to any situations in which waves can be guided without cutoff, including 2D and three-dimensional (3D) acoustic waves and 2D electromagnetic waves. Fig. 1b shows a unit for a 2D acoustic space-coiling metamaterial (of $a$=2.33cm), fabricated by computer numerical control (CNC) milling on aluminum. Fig. 1c shows the full-wave simulation (COMSOL Multiphysics) of an acoustic pressure field for a plane wave at 2.7kHz (or a free-space wavelength of 12.7cm) impinging from the left. The channels (of width 1.5 mm and walls of width 0.8 mm, height is 1 cm) guide the acoustic waves to propagate in a curled fashion. From one corner to the center of the unit, the elapsed phase is found to be delayed by a factor of $n_{1D} = 4.8$, compared with the case when the two end points are connected by a straight line. In another case of electromagnetic waves of H-polarization propagating in 2D, Fig. 1d shows a unit of the metamaterial (of $a$=0.76cm) designed for microwaves. The acoustic hard walls become the copper-wire (of width 0.15 mm and air-channel width of 0.62 mm) layers printed on circuit boards (FR-4 PCB laminate). Identical patterns are then stacked layer-by-layer separated by foam spacers of 2.3 mm thickness. Full-wave simulation (performed at 6.54GHz) shows a similar wave confinement of the H-field in the curled channels (Fig. 1e).

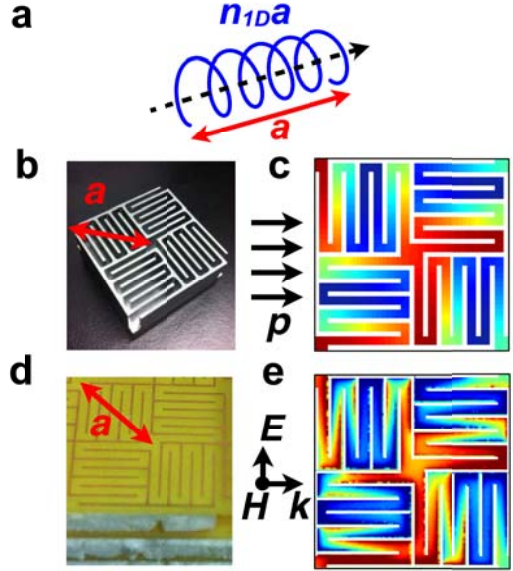

FIG. 1 Principle in coiling up the acoustic and the electromagnetic space. a, Waves confined along a curled channel have an equivalent refractive index $n_{1D} \gg 1$ comparing to propagation along a straight line. b, (photograph) A unit of the acoustic space-coiling metamaterial fabricated using aluminum with $a$=2.33cm. c, Simulated pressure field in a single unit being excited by a plane wave propagating to the right at 2.7kHz, with red/blue color denoting high/low values. d, (photograph) A unit of the electromagnetic space-coiling metamaterial by stacking copper-wire patterns printed on circuit boards, with $a$=0.76cm and foam spacers in the vertical direction. e, Simulated H-field pattern at the copper wire layer being excited by a plane wave propagating to the right at 6.54GHz. White lines represent the aluminum walls or the copper wires.

In the first step, we experimentally obtained the effective medium of space-coiling metamaterials. For an acoustic space-coiling metamaterial, we measured the complex transmission and reflection coefficients at normal incidence by inserting the metamaterial between two waveguide tubes. The effective density ($\rho$), bulk modulus ($B$) and refractive index ($n$) were then extracted using the S-parameter retrieval method [39]. The effective refractive index $n$ from 2.7 to 3.7 kHz varied continuously from −1 to 1 (circles in Fig. 2a). The effective density $\rho$ and reciprocal bulk modulus $1/B$ changed sign at around the same frequency (3.12kHz; circles in Figs. 2b and 2c), corresponding to a normalized frequency $\Omega_0 = \omega a / 2\pi c = 1/n_{1D}$, in which the bands fold back to the zone center [38]. We also extracted the effective medium parameters from full-wave simulations with microstructures included using the same method (Fig. 2 solid lines), which showed very good agreement with the experimental results. Double negativity occurs below $\Omega_0$, while the dispersion around $\Omega_0$ is conical in shape (upper inset of Fig. 2) with a refractive index that can be analytically expressed by $n \approx \sqrt{2}(n_{1D} - 2\pi c/\omega a)$, giving rise to a nearly constant group index $\sqrt{2}n_{1D}$.

While there have been previous demonstrations/ proposals of conical (Dirac-cone) dispersion in the intermediate frequency regime [35-37], our geometric route further relaxes the binding between $\lambda$ and $a$ and makes possible a conical dispersive metamaterial. Indeed, the conical dispersion presented herein is highly isotropic and symmetric in frequency. The group index of the dispersion found from the simulation (blue line, agreeing with the open circles obtained from the experiment) deviates from the constant $\sqrt{2}n_{1D}$ by only 5% in any direction, even at a working frequency of a 15% increase/decrease from $\Omega_0$. In fact, a tiny gap opens up at the cone center frequency because of the finite width of the air channel in constructing the unit cell [38]. However, the size of the gap is less than 1% of the cone center frequency. The analytic formula of the refractive index (a conical description) is thus an accurate representation of the dispersion, and agrees well with the experimental dispersion (open circles in the upper inset of Fig. 2).

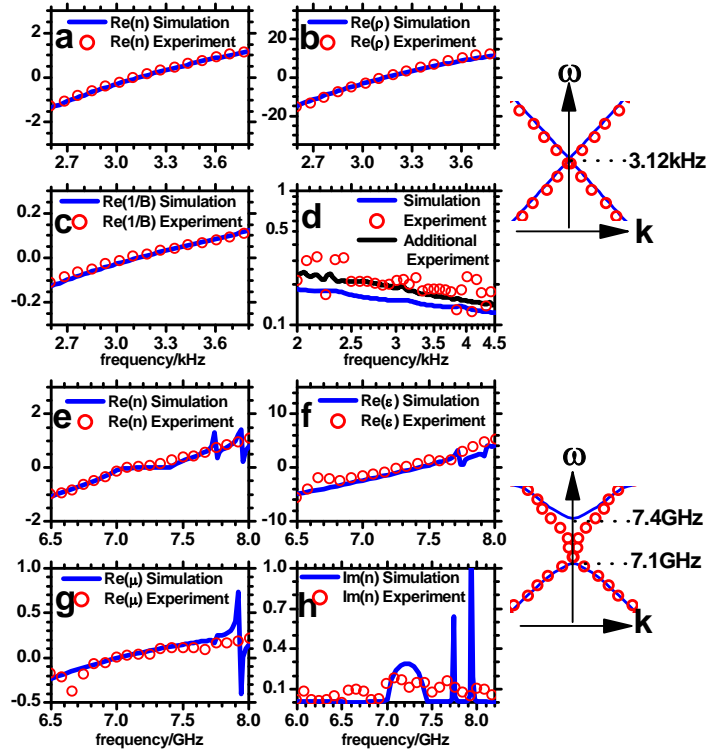

FIG. 2 Effective medium of the space-coiling metamaterials. a-d, The effective parameters of acoustic space-coiling metamaterial: the real part of effective index, density, reciprocal bulk modulus, and the imaginary part of effective index. e-h, The effective parameters of electromagnetic space-coiling metamaterial: the real part of effective index, permittivity, permeability, and the imaginary part of effective index. Upper/lower inset: dispersion for the acoustic/ electromagnetic metamaterial. The blue solid lines/open circles represent the simulation/experimental results.

Although our scheme is geometric in nature in order to avoid resonant absorption, there is still some practical loss. We thus experimentally extracted the imaginary part of the index (Im(n), circles in Fig. 2d). This part agrees with the full-wave simulation results (blue line in Fig. 2d) and is further confirmed by the smoother spectrum obtained from directly measuring the attenuation by cascading a variable number of unit cells (Fig. 2d, black solid line). This absorption loss mainly comes from the viscosity of air within a thin air layer of skin depth $\delta \propto 1/\sqrt{\omega}$ on the inner surfaces of the channels [40]. Moreover, it provides a frequency dependence of $\text{Im}(n) \propto \delta/a \propto 1/\sqrt{\omega}/a$, which agrees with both the experimental and the simulation results (blue line). The current acoustic sample has an experimental figure of merit (FOM=|Re(n)|/Im(n)) of 5.2 at 2.7kHz with $n = -1$. If the size of the metamaterial is scaled up by $s$, the FOM can be further increased by $\sqrt{s}$ times for a fixed value of $n$ (by working at the same normalized frequency in our scheme).

Next, we applied the space-coiling principle to electromagnetic waves. We measured the S-parameters by putting our electromagnetic (EM) space-coiling metamaterial (Fig. 1d) between two standard horn antennas for the H-polarization at normal incidence. Fig. 2e shows the experimentally extracted effective index (circles), together with the index obtained from the full-wave simulations (blue lines) with good agreement, varying from −1 to 1 from 6.54 to 7.95 GHz, again indicating the occurrence of band folding at a very low frequency. Simultaneously negative ε and µ were also found in the negative index regime (Figs. 2f and 2g). From the acoustic to the EM sample, the acoustic hard walls were converted into stratified layers of copper wires of the same pattern, separated by foam spacers of a subwavelength thickness 2.3 mm. The waves were then guided as TEM modes between two adjacent copper wires on the same layer. The H-polarization waves outside the metamaterial can then be coupled to the curled channels for propagation on each layer.

Although the same principle induces double negativity/positivity regimes here again, there is a departure from the ideal conical dispersion in the electromagnetic dispersion (lower inset of Fig. 2). A spectral gap now opens up from 7.1 to 7.4 GHz (a distinct phase of Re(n) ~ 0 in Fig. 2e or between the two zero-crossing frequencies for Re(ε) and Re(µ) in Figs. 2f and 2g). The foam spacer layers with finite thickness splits apart the effective electric and magnetic plasma frequencies. This gives us an additional way for tuning the conical dispersion with a variable gap size. By contrast, decreasing the thickness of the foam spacers would diminish the spectral gap (but this would also require more copper layers to construct the metamaterial as a compromise). In fact, the finite thickness of the foam spacer also introduces spatial dispersion into the flat band of electric plasma ($\varepsilon \sim 0$). Together with the mirror symmetry breaking of the wire pattern, the flat band can be excited in principle and additional sharp resonances are induced to the theoretical effective medium (Figs. 2f and 2g). These sharp resonances are not significant here since they are smeared out by the tiny disorder of the sample, e.g. misalignment between the layers, in the experimental results. Thus, we have achieved a double-negative microwave metamaterial (isotropic in 2D) with a small lattice constant around $a < \lambda/6$, while the experimentally achieved FOM is around 10 (Figs. 2e and 2h at n = −1), including the additional scattering noise owing to the open-space environment (a fluctuating spectrum in the experimental Im(n) results). One advantage of our geometric approach in coiling up space is that the ratio $\lambda/a$ can be easily tuned. $\lambda/a$ can be further increased when we employ a larger number of turns in the wire pattern. For example, if we fix the working frequency for a target refractive index, e.g. −1, a larger number of turns results in a smaller lattice constant $a$ and vice versa. This provides a potential method to make metamaterials work at the deep subwavelength limits, as natural materials do, which is important if we want to manipulate waves with deep subwavelength resolution.

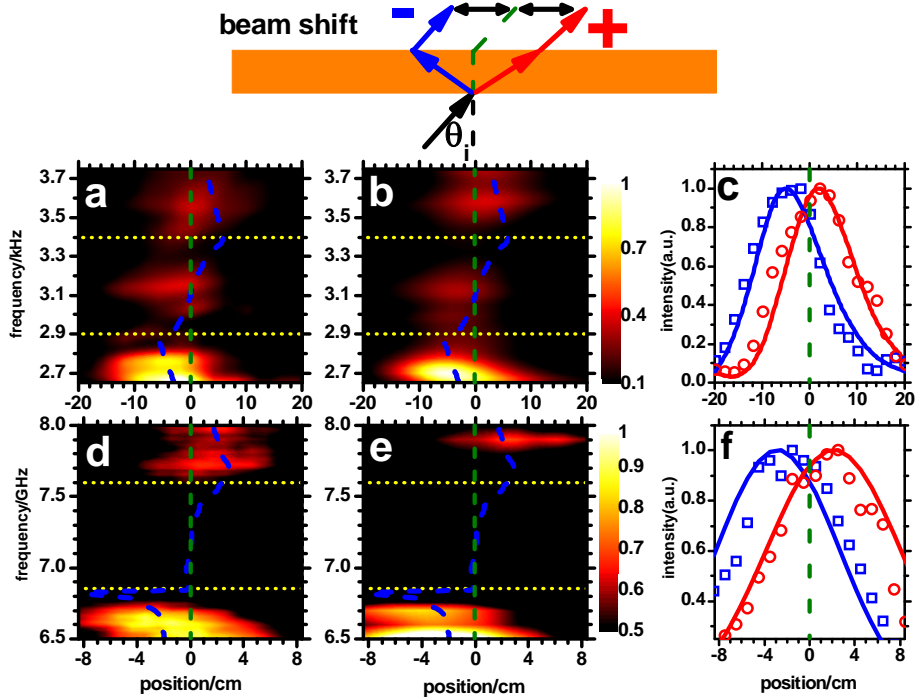

FIG. 3 Beam shift of a slab of space-coiling metamaterial. *inset*, schematic of beam shift experiment. A beam with incident angle $\theta_i = 30^o$ impinges to the slab. The transmitted beam position induced by negative indices locates to the left side of green dashed lines, and vice versa. Intensity profiles of output beam varying with frequency are measured. a Experimental and b simulation results for the acoustic slab of 2x20 units. c, Selected normalized intensity profiles where n=−1 at 2.7kHz (blue) and n=1 at 3.7kHz (red), solid lines represent the simulation results, symbols represent the experimental results. d and e, The electromagnetic cases corresponding to a and b, the EM slab has 3x46 units x 68 copper layers. f, Selected normalized intensity profiles of EM case where n=−1 at 6.54GHz (blue) and n=1 at 7.95GHz (red). Blue dashed lines in a, b, d, e represent the beam center obtained by Snell's law with complex refractive indices.

As the conical dispersion here is isotropic in 2D, the effective medium parameters experimentally established for the space-coiling metamaterials at normal incidence are generally valid for oblique incidence. By constructing the unit cells into a slab, the negative/positive indices below/above the cone center frequency can be revealed by a negative/positive beam shift across a slab. The upper inset of Fig. 3 shows the schematic of this beam shift experiment. A beam with an incident angle $\theta_i$ impinges to the slab and the green dashed line separates the two regions of negative and positive refractions. For acoustics, we construct and put a slab with 2×20 units (6.6 × 66 cm) in the middle of a 2D waveguide. An acoustic beam of 20 cm in width impinges the slab at $\theta_i = 30^o$. The pressure field profile (20cm away from the slab) is then scanned by a microphone (see Figs. 3a/3b for the measured/simulated results) from 2.65 to 3.75 kHz. These two results actually agree very well in terms of beam positions and intensities. From the results, we can divide the frequencies into three regimes by two frequencies (the two yellow dotted lines) having $n = \pm 0.5$, indicating the onset of total internal reflection. From 2.65 to 2.89kHz, the beam center appears on the left-hand side of the vertical green line. This indicates a negative index for our metamaterial. On the contrary, the beam center appears on the right-hand side for positive indices from 3.4 to 3.75 kHz. In particular, the beam intensity profiles at 2.7 and 3.7 kHz where $n = -1$ and 1 are shown in Fig. 3c, confirming the agreement between theory and experiment at both negative and positive refractions. In fact, the absorption inside the slab dampens the effect of multiple reflections, making the beam position generally follow Snell's law employing effective complex refractive indices (blue dashed line) while the intensity fluctuation at different frequencies persists (in both experiment and simulation) because of the impedance mismatch between air and metamaterial. From 2.89 to 3.4 kHz, frustrated total internal reflection occurs. The beam tunnels through the barrier with a very small intensity except near the cone center. In this regime, the transmission coefficient can be approximated as $t \approx \left(1 - r_s^2\right)\exp\left(-|k_z|d\right)$, where $r_s$ is the reflection coefficient of a single metamaterial surface, $k_z$ is the propagation constant in the direction normal to the slab and $d$ is the thickness of the slab. Although the waves generally decay with an exponential factor, there is a resonance in the $r_s$ spectrum across the cone center frequency (owing to zero density there) and a small transmission around the cone center frequency results.

Similarly, the beam-shift experiment for the electromagnetic space-coiling metamaterial, which spans 3×46 units in area with a height of 68 layers of copper wire patterns, is performed. A horn shines a beam (of width 9cm) of H-polarization at $\theta_i = 30^o$, while another horn scans the output beam profile at 16cm away from the slab. The experimental result (Fig. 3d) agrees quite well with the simulation result with an effective medium extracted by the experimental process (Fig. 3e). There are some minor differences in the intensities (at ~ 7.7GHz) because the experiment (for 2D wave propagation) was conducted in a 3D open space instead of a 2D waveguide as in the acoustic case. We can again divide the frequencies into three regimes, namely negative refraction (6.5 to 6.82GHz), frustrated total internal reflection (6.82 to 7.62 GHz) and positive refraction (7.62 to 8 GHz). The experimental result still follows Snell's law quite well. Specifically, the beam profile (blue) at 6.54 GHz where $n = -1$ highlights the negative refraction in Fig. 3f (theory and experiment) in contrast to the positive refraction at 7.95 GHz ($n = +1$). In the tunneling regime, the resonance owing to the conical dispersion disappears because the waves exponentially decay more severely with a larger slab thickness in terms of wavelength in this case.

The isotropic effective medium is also valid for other geometries. We assembled acoustic unit cells into a 45º prism with a size of 4×4 units (Fig. 4a). A beam with width 13.2 cm was then impinged normally on one surface and the refraction angle measured at a radius of 15 cm. From 2.7 to 3.7 kHz, the direction of the experimentally measured refracted beam rotates in an anticlockwise direction, showing a continuous variation from a negative to a positive index (Fig. 4b). This is in good agreement with the simulation results shown in Fig. 4c, confirming the negative index of the sample as well as the validity of our experimentally extracted effective medium.

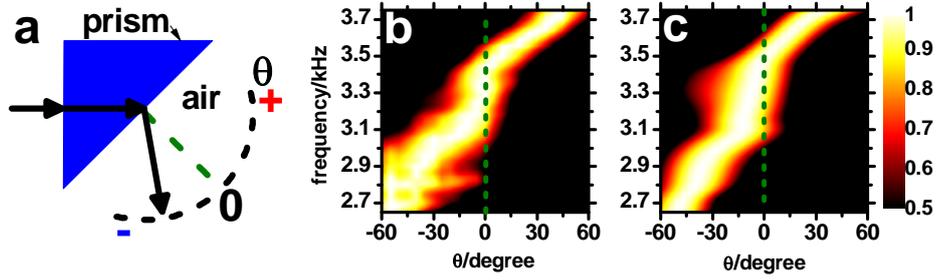

FIG. 4 Prism with acoustic space-coiling metamaterials. a, A metamaterial prism with edge of 45 degrees inclination is impinged by a beam from the left-hand side. The pressure profile of the output beam from the prism is measured at an arc of radius 15cm from −60° to 60° ( 0° is defined along the green dashed lines). b and c, The experimental and simulated pressure intensity profile, normalized to the peak intensity at each frequency.

It is interesting to note other approaches to obtaining a conical dispersion [35-37] in the intermediate frequency regime. Since our geometric approach reduces the first band gap to a very low frequency, it makes a conical dispersive metamaterial possible to construct. This also results in a conical dispersion that is highly isotropic and symmetric in frequency. The group index fluctuates less than 5% even with a 15% increase/decrease in frequency from the cone center frequency. Actually, the ratio of the wavelength to lattice constant can be easily tuned in our geometric approach. In conventional metamaterials that rely on the local resonances of subwavelength structural units, the size of these structural units has to be a certain fraction of the wavelength to support resonances. By contrast, $\lambda/a$ here can be made either smaller or larger by changing the number of turns in coiling up the space. By using the microwave space-coiling metamaterial as an example, and by adding four more turns for the subwavelength channels in each unit, the ratio $\lambda/a$ (at a fixed frequency with $n = -1$) increases from 6 to around 8.1 (35% increase).

In summary, we have demonstrated a geometric approach through coiling up space in order to achieve an acoustic metamaterial with the conical dispersion of constant group velocity and associated double negativity. The presented approach enables us the direct manifestation of negative refraction for isotropic acoustic metamaterials with air-borne sound. The same geometric principle is also valid for designing a double-negative microwave metamaterial where cutoff-free waveguiding can occur, while the ratio between wavelength and lattice constant can be easily tuned. These investigations will prove to be useful for designing future metamaterials with extreme dispersion. For example, a metamaterial with conical dispersion can be used for controlling slow wave propagation in a broadband of frequencies with constant group velocity. By having a metamaterial with wavelength much larger than the lattice constant, we can manipulate waves in a finer resolution and the extreme effective parameters will be useful for employment in the framework of transformation acoustics or optics.

Acknowledgements This work is supported by the GRO program of Samsung Advanced Institute of Technology and the Research Grants Council of Hong Kong (Project No. HKUST2/CRF/11G).